**Temperature increases and thermoplastic microstructural evolution in adiabatic shear bands in a high-strength and high-toughness 10 wt.% Ni steel**

Sung-Il Baik[1,3], Ratnesh K. Gupta[2,4], K. Sharvan Kumar[2,*], David N Seidman[1,3*]

[1.] Department of Materials Science and Engineering, Northwestern University, Evanston, IL 60208, USA.

[2] School of Engineering, Brown University, Providence, RI 02912, USA

[3.] Northwestern University Center for Atom Probe Tomography (NUCAPT), Evanston, IL 60208, USA

[4.] currently at Department of Materials and Metallurgical Engineering National Institute of Foundry and Forge Technology Ranchi-834003 India

* corresponding authors, email: sharvan_kumar@brown.edu (K. Sharvan Kumar); d-seidman@northwestern.edu (David N Seidman)

*Abstract*

A 10 wt.% nickel-steel has been developed for high pressures and low-temperature applications, due to its high strength, excellent toughness, and low ductile-to-brittle transition temperature (DBTT). Under dynamic loading conditions this steel is, however, prone to shear localization that manifests as adiabatic shear bands (ASBs). The temperature increases and thermoplastic microstructural evolution in the ASB are studied in detail, from the macroscopic length scale to the atomic-scale employing correlative electron-backscatter diffraction (EBSD), transmission electron microscopy (TEM), and atom-probe tomography (APT). From a calculation of the temperature increase under adiabatic conditions, based on the conversion of plastic-work to heat generation, the microstructural transitions in the ASB are discussed specifically for: (i) a b.c.c.-f.c.c. phase-transformation and their elemental partitioning; (ii) the thermodynamic model for the compositional change of the V(Nb)-rich carbonitride precipitates during a temperature increase; and (iii) grain-refinement and rotation by dynamic/mechanical recrystallization processes. Solute segregation at subgrain boundaries, measured using the Gibbsian interfacial excess methodology, reveals how solute segregation contributes to the instability of localized shear-deformation by promoting the depinning of solute elements and the migration of a grain boundary. Finally, a kinetic model for grain refinement/rotation within an ASB is described by the dynamic recrystallization behavior with: (i) subgrain formation; (ii) rotation/refinement by deformation; and (iii) grain growth by subgrain coalescence with further rotation and a temperature increase. The temperature increase under dynamic deformation in an ASB promotes grain boundary migration and subgrain coalescence to create a large degree of equiaxed grains with a low density of imperfections.

1. Introduction

Structural materials with high toughnesses and high strengths have attracted much attention recently to satisfy the increasing demands for safety and reliability under extreme working conditions. Due to their high strength and excellent crack-arrest toughness, high-nickel steels are promising candidates in earthquake-resistant structures, blast-resistant Naval ship structural components, and for liquefied natural gas (LNG) tankers [1-4]. Low-carbon concentration (<0.1 wt% C) in steel is suitable for Naval structural applications for good weldability [5, 6] and a Ni-rich austenite-phase formation in a tempered martensitic matrix increases ductility and toughness in steels [7, 8]. Specific heat treatments, such as the quenching-lamellarizing-tempering (QLT) procedure [9-12] enhance the strength and toughness of 10 wt.% Ni steels for use at cryogenic temperatures by stabilizing the austenite-phase (up to ~20 vol.%) in a tempered martensitic matrix. The stable austenite reversion from a martensitic matrix by an intercritical annealing step improves the low-temperature Charpy impact toughness of 153 J at -84.4 °C [13], by lowering the ductile-to-brittle transition temperature (DBTT), thereby improving its performance that would otherwise be adversely affected by the inherent brittleness of the hard martensitic-phase [14-16]. Despite the excellent static mechanical properties of these high Ni-steels, there is an acute propensity for localized deformation during high strain-rate deformation [17].

A thermoplastic-instability resulting in localized plastic flow arises in high strain-rate tests as a consequence of localized temperature increases and is a major limitation for many safety-critical structural materials [18, 19]. An adiabatic shear-band (ASB) is a significant material failure mode resulting from the localized plastic flow in dynamic-loading conditions, such as metal forming, machining and ballistic impact [20-23]. The term *adiabatic* is a thermodynamic term meaning an absence of heat transfer because heating is locally faster than the time scale at which heat is conducted away from a plastically deforming region. The lack of time for thermal diffusion results in the suppression of work-hardening and the occurrence of localized softening, leading to material failure [24, 25]. This adiabatic shear localization phenomenon is observed more readily in materials with low strain-rate sensitivity, a small thermal conductivity, and a high thermal softening rate [26, 27]. Experimental analyses of ASBs have demonstrated that microstructural changes due to high strain-rate deformation resulting in strain localization (large local plastic strains and high strain rates) and local temperature-increases can include dynamic recovery, recrystallization, phase transformations, melting, and/or amorphization [28, 29]. Underlying mechanisms responsible for these microstructural manifestations within an ASB are complex and are still not fully understood because they are caused by combined thermal and mechanical changes occurring in very short durations of the order of microseconds to milliseconds in response to an external load.

Since the direct measurement of temperature within localized ASBs is difficult given the microseconds duration, indirect techniques including infrared detectors have been used to estimate the temperature increase; however, their spatial and time resolutions are limited to the micrometer length scale [30]. All the plastic strain is highly concentrated when the shear band region is severely plastically deformed, leading to a drastic change in structure and dilatation in a short period of time. The local temperature increase during high strain-rate deformation can influence the local microstructural evolution, including precipitation and dynamic/mechanical recrystallization within a shear band [18, 31]. In the case of a low-carbon steel, a temperature increase within the ASB can produce austenite (f.c.c.) phases, followed by its conversion to martensite during subsequent rapid cooling [32, 33], and carbide precipitates can dissolve or reprecipitate during local temperature increases [34, 35]. Solute atoms may also segregate at dislocations and grain boundaries (GBs), and the supersaturation of solute atoms may be sufficiently high to nucleate precipitates during subsequent cooling. However, the various microstructural aspects during shear localization and their underlying mechanisms are not fully understood yet in the 10wt.% Ni steel.

Herein, we present a detailed description of microstructural evolution and provide a thermoplastic model of adiabatic shear bands in a 10 wt.% Ni steel, utilizing correlative electron backscatter diffraction (EBSD), transmission electron microscopy (TEM), and atom-probe tomography (APT) experiments. The microstructural transitions occurring in the shear band during the short times over which local temperature increases due to adiabatic heating are discussed with respect to phase transformation, precipitate formation, elemental GB segregation, grain refinement, and growth through dynamic recrystallization processes. These experimental results coupled with analytical modeling of the thermoplastic processes that prevail, provide a comprehensive understanding of the microstructural evolution in a high-nickel steel during dynamic deformation.

## 2. Experimental details

The material examined in this study has the nominal composition, 0.32C-0.04N-0.61Mn-0.41Si-0.13Cu-0.64Cr-0.64Mo-0.08V-0.005Nb-9.75Ni-Balance Fe (at.%). In weight percent, the alloy composition is 0.07C-0.01N-0.60Mn-0.21Si-0.15Cu-0.60Cr-1.11Mo-0.07V-0.009Nb-10.2Ni-Balance Fe. A sample was austenitized at 843 °C for 1 h, water quenched, and then tempered at 454 °C for 5 h, followed by water quenching to room temperature. Next, a quasi-static compression test was performed on a cuboidal specimen (4.2x4.2x6.3 mm$^3$) at a strain-rate of $10^{-4}$ s$^{-1}$. Dynamic compression tests employing a split-Hopkinson bar set-up were conducted at strain-rates ranging from 1000 s$^{-1}$ to 3000 s$^{-1}$, and metallographic observations of a deformed specimen that was tested at 2800 s$^{-1}$ confirmed shear localization through the presence of an ASB with a width of ~20 μm.

TEM and APT samples were prepared by lifting-out specimens, utilizing Ga$^+$ ion-milling employing a FEI Helios dual-beam focused ion beam (FIB) microscope, utilizing a final energy of 2 kV at 24 pA. The TEM sample was lifted out from selected regions in the ASB and then transferred to a Cu grid [36, 37] for characterization by EBSD, TEM, and APT. The scanning electron microscopy (SEM) and EBSD analyses utilized a field-emission-gun, scanning electron-microscope (FEI Quanta 600F) equipped with an HKL Nordlys S camera. The accelerating voltage and working distance were 15 kV and 10 mm, respectively, with the sample stage tilted by 70° with respect to the specimen surface normal. The microstructure and crystallography were characterized by a JEOL 2100F TEM operating at 200 kV. The selected area diffraction patterns (SADPs) were indexed using the programs *Crystal-Maker* and *Single-Crystal* [38]. APT nanotips were fabricated from a TEM sample to isolate a specific region of interest (ROI) within the ASB [39, 40].

A Cameca LEAP 4000X-Si tomograph was employed to measure the compositions of APT nanotip samples. Picosecond pulses of ultraviolet (UV) laser-light (wavelength = 355 nm) were utilized to evaporate individual atoms at a pulse repetition rate of 250 kHz, a laser energy of 30 pJ pulse$^{-1}$, and an average detection rate of 0.01 ions pulse$^{-1}$. The nanotip temperature was maintained at 60 K, and the gauge pressure of the analysis chamber was < 2.7×10$^{-9}$ Pa (2×10$^{-11}$ Torr). Data analysis was performed on 3-D reconstructions, utilizing Cameca's reconstruction program IVAS 3.6.1, and compositional information was obtained employing the proximity histogram methodology [41].

## 3. Results

### 3.1 The strain-stress curve

Two compression stress-strain curves, one quasi-static (obtained at a nominal strain rate of $10^{-4}$ s$^{-1}$) and one dynamic (produced at a strain rate of 2800 s$^{-1}$ utilizing a split-Hopkinson bar set-up), are compared in Fig. 1. The quasi-static yield stress is ~1.1 GPa and this is followed by mild work hardening to about of 1.25 GPa after ~18% plastic strain. It is noted that the stress-strain curves shown in Fig. 1 are true stress-strain curves, so that the mild work-hardening mentioned above is real and not due to the progressive increase in the specimen cross-section area with increasing compression strain. In contrast, dynamic deformation results in an increased average initial flow stress of ~1.4 GPa. The serrated appearance of the dynamic stress-strain curve, where the stress amplitude slowly decays at larger strains, is a phenomenon often observed in the Split-Hopkinson bar experimental output and is associated with wave dispersion. In the first ~10% of plastic deformation, the work-hardening rate is near zero following after which *work softening* is observed. Towards the very end of the curve, a sharp decrease in the stress level is likely related to the end of the pulse duration in the specimen. The deformation evolution during dynamic deformation observed can be described as occurring in two stages. A first stage representing more-or-less homogeneous deformation followed by a second stage, where the localization commences. As deformation progresses, the magnitude of the localized strain and extent of localization becomes more acute and an adiabatic shear band evolves. The microstructural evolution in this a highly deformed localized volume is investigated by correlative SEM, EBSD, TEM, and APT experiments.

### 3.2 Site-specific adiabatic shear band microstructure analyses

#### 3.2.1 Scanning electron microscope results

Fig. 2 displays an SEM image of shear localization in a well-defined ASB. The specimen was mechanically polished and etched with a 2 % Nital solution. In Fig. 2(a), the structure flanking the ASB is lath martensite [42] while the interior of the ASB, based on grain shape, is comprised of two distinct regions. Plastic flow in the ASB is discerned by the closely-spaced striations aligned in the shear direction. The striations are denser towards the core of the ASB and are more dispersed as one moves towards its edges; these regions on either side of the central region are marked as a *deformed shear band* and have a width of ~21.7 ± 1.5 μm. In the middle of the ASB there is an approximately 3.3 ± 0.5 μm thick localized area with a gray contrast, marked as a *transformed shear band*. Fine microcracks are present within the transformed shear band and their linkage may lead to fracture [30, 43]; an example of one such microcrack is displayed in Fig. 2(b). A plan-view TEM/EBSD specimen was obtained from the location just ahead of this crack in the ASB region as illustrated in Fig. 2(b).

### 3.2.2 Transmission electron microscopy results

TEM was employed to examine the microstructure within the ASB at a fine scale. Fig. 3(a) displays a low-magnification TEM image with the foil plane parallel to the top surface of the specimen in Fig. 2(b) with the entire width of an ASB, including a crack site. The locations identified by the rectangles and marked *"b"* and *"c"* were analyzed using bright-field (BF) and dark-field (DF) microscopy with selected area diffraction patterns (SADPs). The DF images, Figs. 3(e, f), were obtained from the (110) reflections indicated by white circles in the SADP in the inset. The microstructure in the as-quenched sample has been previously shown to be lath martensite consisting of a high dislocation density [44]. The size distribution of martensite lath widths revealed that the laths in the deformed specimen were significantly refined and well-aligned in the shear direction, with respect to those in the as-quenched sample. In this study, the TEM specimen was obtained from within the severely deformed regions of the ASB. BF and DF TEM images indicate clearly the change in grain shape within the shear band, from an elongated lamellar morphology (corresponding to *the deformed shear band*) to a highly misoriented equiaxed structure towards the center of the shear-band (corresponding to the *transformed shear-band*). This transition is marked by a white dashed line in Fig. 3(b). The diameter of the recrystallized equiaxed grain increased from ~100 nm in the vicinity of the transition line to a value of ~200 nm in the core of the ASB. The changes in orientation and dimensions are also reflected in the SADPs. The discrete nature of the SADP means that fewer grains are involved in the selected area aperture (~10 μm) and grains have a (110)-preferred orientation with respect to the shear-plane normal (SPN) direction, Fig. 3(e). This tendency changed to a well-distributed ring pattern in the transformed shear band, Fig. 3(f), which represents a microstructural transition from large-sized textured grains to smaller and more randomly oriented grains. Additionally, a microcrack is also observed ahead of the main crack and it appears to be propagating along the grain boundary, Fig. 3(d).

### 3.2.3 Electron back-scatter diffraction (EBSD) results

High-resolution electron back scatter diffraction (EBSD), Fig. 4, was employed to characterize the fine-grained structure in the transition region from the deformed to the transformed band of the ASB. The EBSD area corresponds to the rectangle labeled "*EBSD*" in the TEM image in Fig. 3(a). The orientation map across the transition line of the ASB is revealed by an inverse pole figure (IPF) within a reference frame: a shear-direction (SD, X-axis) and shear-plane normal (SPN, Y-axis. In this image, the grain boundaries are represented by black lines and identify a misorientation greater than 15 °, while small misorientations (<1.5 °) are identified by gray lines and twin boundaries by white lines ( <111>, 60º, {112}). The transition from elongated lamellar microstructure to equiaxed small grains is represented by a blue dashed transition line (TL). Above the TL, the color-gradient within the elongated lamellar structure indicates a lattice distortion within a range of 1º - 2º. This distortion is accommodated through a dislocation cell structure and a subgrain boundary (sub-GB). The fragmented grains below the TL share a similar color code on the map, implying that they are derived from the elongated lamellar grains in the deformation band by further bending or splitting. A shear texture ({101}, <111>) is also observed in the pole figure map, Fig. 4(b). This implies that <111>-direction of the b.c.c. crystal lattice is parallel to the shear-direction (X-axis) and the normal to the {110} plane is parallel to the shear-plane normal (Y-axis) in agreement with the SADP analysis of the elongated lamellar structure, Fig. 3(e). This preferred orientation disappears in the transformed shear band, where the microstructure is composed of equiaxed grains;  this aspect is discussed in detail later in the paper.

## 3.3 Correlative Transmission electron microscopy and atom probe tomography analyses

### 3.3.1 Deformed shear-band

Fig. 5 displays a 3-D APT analysis of a specimen obtained from the deformed shear band that was located ~ 3 μm away from the TL. The correlative TEM analysis of an APT nanotip, Fig. 5(a), presents the diffraction contrast in each grain. The SADP in the inset was obtained from the grain between GB-3 and GB-4 and is indexed as the [112] zone-axis of α (b.c.c.) iron. Figs. 5 (b,c) display two 3-D APT reconstructions with ~133 and ~120 million atoms, respectively. The second APT sample (red rectangle in (a)) was obtained by re-sharpening a blunted APT nanotip, utilizing the dual-beam FIB microscope's $Ga^+$ ion beam, after the first APT experiment (blue rectangle in (a)). The elements are color-coded in the APT reconstructed images with red dots (Fe), dark green dots (both N and Si), and black dots (C). It is difficult to distinguish between the N and Si peaks in an APT experiment because the peaks overlap in the time-of-flight (TOF) mass-to-charge state-ratio spectrum at 14 amu, respectively. The GBs in the 3-D reconstructed APT images are identified by comparing the segregation behavior of elements with the GB locations in the reference TEM image. Inside the grain, the sub-GBs are detected by N (or Si) segregation (dark green dots), which are produced by solute segregation at the dislocations and GBs, possibly during the shear deformation. This N-segregation at a sub-GB or GB was not observed in the undeformed heat treated 10 wt.% Ni steel [45]. The level of segregation at an interface is represented by concentration profiles, Figs. 5(d,e). The concentration profiles contain the elements Fe, Ni, C, N or Si, Mn, and H in high angle GBs, Fig. 5(d), and in sub-GBs, Fig. 5(e). The segregated element detected at a low angle sub-GB is nitrogen due to its fast diffusion in a b.c.c. matrix and its strong interactions with dislocations in this sub-GB.

The equilibrium segregation of a solute species, $i$, at a GB is described by the Gibbsian interfacial excess [46], $\Gamma_i$, at constant temperature and pressure, which is defined by the excess number of solute atoms $i$, $N_i^{excess}$ per unit area, A. The Gibbsian interfacial excess, $\Gamma_i$, is determined from the following equation [47]:

$$\Gamma_i = N_i / A = \rho \Delta x \sum_{m=1}^{p} (c_i^m - c_i^o) \qquad (1)$$

where $\rho$ is the atomic density (85 atoms·nm$^{-3}$ for b.c.c. iron); $\Delta x$ is the distance between the data points (0.5 nm); $c_i^m$ and $c_i^o$ are measured and average concentrations of an element $i$ in the matrix, respectively. The concentration in the matrix, maximum concentrations at the GB and sub-GB in atomic percent (at.%), and their $\Gamma_i$ (atoms/nm$^2$) values are listed in Table 1. The segregation behavior of each GB is different because it depends on an individual GB's characteristics and interactions of each element with a given GB [48-50]. The highest concentration position at GB-2 and sub-GB-4 are selected to measure the interfacial excess, $\Gamma_i$. The carbon concentration is 1.07 at. % at GB-2, corresponding to an enhancement factor of 4 with respect to the ferrite (b.c.c.) matrix. There is also an increase of N or Si to 1.47 at.%, which is an enhancement factor of 3. Hydrogen was also detected in this sample at a value of 0.90 at.% at GB-2, an enhancement factor of 2 with respect to the ferrite (b.c.c.) matrix. In contrast, Mn has smaller increase of concentration than that of the ferrite (b.c.c.) matrix. These enhancement factors can also be presented quantitatively as interfacial excesses, $\Gamma_i$, with positive (segregation) and negative values (depletion). Nitrogen, including Si, is strongly segregated at GB-2 with a maximum $\Gamma_{N(Si)}$ of 2.56 ± 0.13 atom·nm$^{-2}$. The elements Ni, C, and H segregate with $\Gamma_i$ values of 1.65 ± 0.43, 1.88 ± 0.12, and 1.07 ± 0.15 atom·nm$^{-2}$, respectively. Manganese is segregated slightly with a $\Gamma_{Mn}$ value of 0.20 ± 0.13 atom·nm$^{-2}$. A sub-GB can be generated within a martensite lath by dislocation rearrangements during the intense shear accompanying localization. Although the local temperature may not reach the austenite phase-transformation temperature, the rearrangement of dislocations produces several sub-GBs inside an original lath. The Gibbsian interfacial excess values of the following elements, N(Si), Ni, C, Mn, and H, are: 2.29 ± 0.15, 1.02 ± 0.58, 0.40 ± 0.12, 0.11 ± 0.15 and 1.08 ± 0.25 atom·nm$^{-2}$, respectively.

### 3.3.2 Transformed shear-band

The shapes of grains and elemental distribution in the core of a shear band differ from those at its periphery, likely related to the adiabatic heating resulting from the intense shear localization (locally high strains and strain rates) and the possible consequential phase-transformations and/or dynamic recrystallization. Fig. 6 presents a correlative TEM and 3-D APT analysis of the transformed shear-band, which is obtained from the core region of the ASB. The BF-TEM image of the nanotip is presented in Fig. 6(a), which contains six grains and a corresponding 3-D APT reconstruction, ~133 million atoms, Fig. 6(b). The APT analysis intersects two *"dark-contrast grains"* that display enrichment of carbon and other austenite stabilizing elements, Ni, Mn, and N, in the concentration profiles, Fig. 6(c). The spikes in the C and N concentrations are related to the austenite/ferrite interface and a nitride precipitate, respectively. This high level of C in austenite was not observed within individual grains in the deformed shear band region, Fig. 5. The presence of fine austenite grains in the ASB in a similar alloy was reported previously and confirmed using microdiffraction [44]. In their TEM-EDS analysis of an ASB, the partitioning behavior of the substitutional elements, Ni and Mn, to the austenite-phase are unclear when compared with the composition of the ferrite-phase [44]. This means that the interstitial elements, particularly carbon, likely play an important role in the ferrite-to-austenite phase transformation by stabilizing the austenite structure in the ASB. This austenite-phase can transform to martensite during cooling; it is very difficult, however, to transform austenite to martensite in the nanometer scale grain size even though the composition may permit $M_s$ to be above room temperature [45]. 3-D APT is capable of measuring concentrations with high-fidelity and a detection limit of ~10 at. ppm for mass spectra with low background noise [51-53] and can therefore resolve the redistribution of solute elements during dynamic recrystallization. The carbon concentration increases up to a maximum value of 1.5 at.%, with the partitioning ratio of carbon between the austenite- and ferrite-phases ($K_c^{\gamma/\alpha}$) is 5.09 ± 0.27. The elements, N(Si), Mn, and Ni are also slightly enriched in the austenite grain with partitioning ratios ($K_i^{\gamma/\alpha}$) of 1.30 ± 0.19, 1.38 ± 0.07, and 1.08 ± 0.19 at.%, respectively. The detailed concentrations of the ferrite-matrix and the austenite-phase are presented in Table 2.

### 3.3.3 Vanadium-rich carbonitride precipitate

Vanadium-rich carbonitride precipitates were also observed in the core of the ASB (transformed band), Fig. 7, where the precipitates are represented by a 20 at.% vanadium iso-concentration surface in the APT reconstruction [41]. This nanoscale precipitation is another important microstructure evolution arising from the local temperature increase within an ASB. When the temperature rises high enough above the austenitic transition-temperature, the relevant carbide-forming elements (V, Nb, Cr etc.) dissolve in the base metal (Fe), and then re-precipitate as nanoscale carbonitrides when the ASB region cools down rapidly. From two APT analyses, Fig. 7(a), the diameter and number density of the V-rich nitride-precipitate were determined to be 3-7 nm and 2.2 ± 0.6 x $10^{23}$/m$^3$, respectively. The chemical compositions of the precipitates were measured in the plateau regions in a proximity-histogram (proxigram) concentration profile, Fig. 7(b). The concentrations of the main constituents in the precipitate are 40.16 ± 1.45 at.% V and 34.51 ± 0.94 at.% N. The elements, Nb, Cr, Mo, and C are also present at up to 7.65 ± 0.76, 4.12 ± 0.30, 0.77 ± 0.25, and 2.67 ± 0.46 at.%, respectively. This elemental enrichment is accompanied by partitioning ratios between the precipitate and ferrite-matrix concentrations ($K_i^{p/\alpha}$). The partitioning ratios of V, Nb, N (Si), and C are 334.7 ± 30.4, 382.5 ± 84.2, 63.91 ± 2.94, and 12.0 ± 0.71, respectively. The nitrogen partitioning is increased by 3450 times with respect to a nominal nitrogen composition of 0.04 at. %. The detailed concentrations in the precipitate and the matrix, and their partitioning ratios are presented in Table 2. Another interesting feature is the core-shell structure of precipitates similar to $Al_3Sc$ precipitates hardened Al-based alloys [54-56]. The concentrations of V and N continuously increase in the core of the precipitates (core structure); Nb, Cr and C increase initially at interfaces but decrease again in the core of the precipitates (shell structure) [57, 58]. This will be discussed later.

## 4. Discussion

### 4.1 Temperature increase estimation under adiabatic conditions

Temperature increases within an ASB play a major role in determining the thermoplastic mechanical behavior resulting during high strain-rate deformation. Because of the experimental difficulty related to the direct measurement of temperature rise within a narrow ASB in a short duration of the order of milliseconds, temperature increases are usually estimated by the conversion of plastic work to heat generation utilizing the following equation [20, 59]:

$$\Delta T = \left(\frac{V_o}{V_p}\right)\frac{\Psi}{\rho c_p}\int_0^t \sigma d\varepsilon \quad (2)$$

where $\Psi$ is the fraction of plastic work converted to heat which is generally taken as ~90% [60], $\rho$ is the density (7.874 g/cm$^3$), $c_p$ is the specific heat capacity (0.450 J/gK), σ is the applied stress, and ε is the strain. A factor, $V_o/V_p$, reflects the ratio of the total deformed volume to the transformed shear-band volume. By assuming that shear localization commences immediately after the onset of plasticity during high-strain rate deformation, a volume ratio of deformed/transformed shear-band ($V_o/V_p$) of ~ 6 is obtained from the SEM image shown in Fig. 2. The stress-strain curve in a compression test was fitted to dislocation-mechanics-based constitutive relations [31, 61]:

$$\sigma = \sigma_o + C_1 \exp[-C_2 T + C_3 T \ln(\dot{\varepsilon})] + C_4 \varepsilon^n \quad (3)$$

where $\sigma_o$, $C_1$, $C_2$, $C_3$, $C_4$, and n are the parameters determined using a nonlinear regression analysis. The dependence of the yield-stress ($\sigma$) on temperature (T) and strain-rate ($\dot{\varepsilon}$) in a b.c.c. metal can be represented by exponential terms of the form, $C_1 \exp[-C_2 T + C_3 T \ln(\dot{\varepsilon})]$. Additionally, a separate plastic strain-hardening contribution to the flow-stress is represented by a power-law expression of the form, $C_4 \varepsilon^n$. The $\sigma_o$ term reflects the theoretical yield stress of a b.c.c. metal plus grain-size strengthening, ~1.1 MPa, obtained from the quasi-static yield stress in Fig. 1. The best-fit parameters for the stress-strain curve are: $C_1 = 570$ MPa, $C_2 = 0.0072$, $C_3 = 0.0007$, $C_4 = 320$ MPa and n = 0.47. Computed isothermal and adiabatic stress-strain curves are displayed in Fig. 8 for two strain-rates, 2800 s$^{-1}$ and 10$^{-4}$ s$^{-1}$, at 300 K. We note that the oscillations observed in the dynamic stress-strain curve were not taken account of in the simulation.

The temperature increases obtained according to eq. (2) and using the stress-strain form in eq. (3) fitted with the above parameters, are plotted in Fig. 9 for multiple scenarios. Under adiabatic conditions, the temperature can reach 924 °C when the global height strain is of the order of 40% (the specimen whose dynamic stress-strain curve is shown in Fig.1 fractured just short of 40% height strain). Again, this estimate (red dotted line in Fig. 9) assumes that shear localization commences at the onset of plasticity during high-strain rate deformation. At the other extreme, if homogeneous deformation is assumed all the way to 40% strain (solid blue line in Fig. 9), the temperature increase is small, ~ 146 °C. The temperature rise could be much higher than the calculated value in the first case if the strain was more localized at the core of the ASB. For example, the green dashed line in Fig. 9 shows the situation when the shear localization commences around 35% height strain but with a high localization factor ($V_o/V_p$) of 50. In this case, the temperature rises quickly and can even approach/exceed the melting temperature of the alloy.

Such a temperature increase can cause local structural changes in the ASB, including the formation of submicron-sized equiaxed grains, austenite formation, carbide dissolution and precipitation, and in the extreme, the formation of a nanometer-scale liquid film due to local melting [44, 62]. The local temperature increase can be heterogeneous within a shear band and the development of *hot-spot regions* in the ASB provides potential sites for crack nucleation [43, 63]. The volume change associated with a phase transformation and the subsequent temperature decrease can also nucleate micro-voids in many locations in the ASB, leading to sample fracture [64]. The addition of high Ni-concentration (10 wt.%) to a low carbon steel expands the austenite phase region and decreases its transformation temperature to 678 ºC [45], which is significantly lower than the estimated temperature of 924 ºC within the ASB, and in principle the b.c.c. ferrite/martensite phase can be fully transformed to the austenitic phase upon adiabatic heating. The kinetics for the ferrite-to-austenite phase transformation may, however, be incomplete and permit only partial austenite-transformation. Therefore, only the interstitial elements, carbon and nitrogen, can diffuse in the austenite phase due to their larger diffusivities in iron. The carbon concentration in the austenite/martensite phase transformation increases from a value of 0.22 ± 0.01 at.% in the ferrite phase to a maximum value of 1.5 at.% C, with an average concentration of 1.12 ± 0.06 at.%, Table 2. In contrast, Mn and Ni are only weakly enriched in the austenite/martensite grain, with concentrations in austenite of 1.05 ± 0.05 and 11.48 ± 0.16 at.%, respectively, compared to their values in ferrite of 0.77 ± 0.02, and 10.59 ± 0.08 at.%, respectively. The small partitioning ratio of N(Si), 1.30, is due to the migration of Si (ferrite-stabilizer) and N (austenite) in opposite directions. The austenite chemistry in the shear band can be compared to that for the austenite phase of a 10 wt.% Ni steel subjected to a QLT heat treatment [65], where the average concentrations of C, N(Si), Mn, and Ni in the austenite phase are reported to be 0.38 ± 0.01, 0.51 ± 0.01, 1.30 ± 0.01, and 13.97± 0.06 at.%, respectively [42]. Carbon and N(Si) have larger partitioning ratios, whereas, Ni and Mn exhibit smaller partitioning ratios in the ASB relative to the QLT heat-treated austenite phase, implying that carbon and nitrogen play major roles as austenite stabilizers, and prevent a reversion to the ferrite-phase when this steel undergoes rapid-cooling after high strain-rate deformation. It is emphasized that the small grain size in the ASB can also prevent the austenite from forming martensite during rapid cooling after deformation [44]. Additionally, the interstitial elements distributed in the austenite phase can also return to a local region during cooling, which was found in many air-cooled dual phase steels [66, 67].

**4.2 V(Nb)-rich carbonitride precipitate behavior in an adiabatic shear-band**

The APT experiment detected V(Nb)-rich carbonitride precipitates in the transformed shear-band at a density of $2.2 \pm 0.6 \times 10^{23}/m^3$ with a diameter range of 3-7 nm. The concentrations of V and N in these precipitates were 40.95 ± 1.07 at.% and 32.89 ± 0.94 at.%, respectively. Niobium and carbon were also detected at concentrations of 6.43 ± 0.37 and 1.66 ± 0.18 at. %, respectively (Table 2). This VN-type precipitate has a NaCl-type crystal structure (f.c.c.), with a lattice parameter of a = 0.4139, but the lack of nitrogen in the VN precipitate results in a nitrogen vacancy ($v$) structure, $V_4N_3$ [68]. Cubic MC-type precipitates (M = V, Ti and Mo) with a mean diameter of 20-25 nm were reported in an as-quenched sample [44]; the number density appeared, however, to be low compared to the current APT observation. The low density of MC carbides in the base metal is also supported by the fact that no precipitates were found in the deformed shear bands from several APT results, Fig. 5. Even if large-sized carbides present in the undeformed material are mechanically broken down by the shear deformation, these small precipitates within the ASB would have still been easily resolved; furthermore, the composition of the carbide is also changed, similar to that observed in the austenite phase transformation in the core of the shear band. The dissolution and compositional change of the precipitates in the 10 wt.% Ni steel are also believed to be a consequence of the temperature increases within the ASB. Similar observations and inferences have been previously made; specifically, the creation of high thermal energy input during intense plastic shearing in a carbide-hardened steel was shown to result in a refined grain structure and nearly complete carbide dissolution in steel [69, 70].

The concentration of carbonitride precipitates in a Nb/V micro-alloyed steel is estimated by a quasi-regular-solution thermodynamic model as a function of temperature and composition [71-73]. The small addition of Nb and V in combination with N and C yields a V(Nb) carbonitride precipitate with the following composition:

$$Nb_xV_{1-x}C_yN_{1-y}, \text{ where } 0 \leq x, y \leq 1 \qquad (4)$$

The equilibrium concentrations of x and y, and the mole-fraction of carbonitride precipitates, $f$, can be predicted by the solubility product of each pure binary compound with the assumption of perfect stoichiometry (total number of Nb plus V atoms is equal to the total number of C and N atoms in the precipitate).

In a dilute solution model [74], the austenite/carbonitride equilibrium can be described by the following three equations, using the solubility product (K) of each pure carbide and nitride [71],

$$y \ln \frac{xy \cdot K_{NbC}}{Nb_e \cdot C_e} + (1-y) \ln \frac{x(1-y) \cdot K_{NbN}}{Nb_e \cdot N_e} = 0 \quad (5a)$$

$$x \ln \frac{xy \cdot K_{NbC}}{Nb_e \cdot C_e} + (1-x) \ln \frac{(1-x)y \cdot K_{VC}}{V_e \cdot C_e} = 0 \quad (5b)$$

$$x \ln \frac{x(1-y) \cdot K_{NbN}}{Nb_e \cdot N_e} + (1-x) \ln \frac{(1-x)(1-y) \cdot K_{VN}}{V_e \cdot N_e} = 0 \quad (5c)$$

where $Nb_e$, $V_e$, $C_e$, and $N_e$ are the equilibrium mole fractions of each species in austenite. The solubility product ($K_s$) in austenite is used to predict the equilibrium phases in a multicomponent alloy employing a thermodynamic database. The solubility products are given in the Henrian form:

$$\log_{10} K_s = \log_{10} [M][X] = = A - B/T \quad (6)$$

where T is temperature in Kelvin and the solubility data are given in weight percent. [M] and [X] are the dissolved V or Nb, and N or C concentrations (wt.%) in the b.c.c. matrix. The solubility parameters A and B in the pure binary compound are determined experimentally [75-78], and are presented in Table 3. Equation 5 can be solved assuming mass balance during the reactions; the total number of atoms of each of the elements is conserved during the reactions and the equilibrium mole fractions are expressed as:

$$Nb_e = \{Nb_s - f(x/2)\}/(1-f) \quad (7a)$$

$$V_e = \{V_s - f[(1-x)/2]\}/(1-f) \quad (7b)$$

$$C_e = \{C_s - f(y/2)\}/(1-f) \quad (7c)$$

$$N_e = \{N_s - f[(1-y)/2]\}/(1-f) \quad (7d)$$

where $Nb_s$, $V_s$, $C_s$, and $N_s$ are the mole fractions of each species in the steel before precipitation occurs and $f$ is the mole fraction of the precipitate at equilibrium. The nominal concentrations of each species in the steel are used for the calculations, and the equilibrium compositions of the precipitate and matrix and the precipitate mole-fraction (x, y, $Nb_e$, $V_e$, $C_e$, $N_e$, and $f$) can be determined for a specified temperature. The values for x, y, and the mole fraction of the carbonitride precipitate are displayed in Fig. 10. The effective temperature for complete dissolution is about 1058 ºC. When the temperature decreases, the carbon fraction in the precipitates increases, and the Nb fraction decreases. At a given estimated temperature, 924 ºC in eq. 2, the composition of the carbonitride is $Nb_{0.17}V_{0.83}C_{0.12}N_{0.88}$ (x = 0.17 and y = 0.12), and the mole fraction ($f$) is $5.3 \times 10^{-4}$. These values are in reasonable agreement with the APT experimental result for the composition of the V(Nb) carbonitride-precipitate ($Nb_{0.16}V_{0.84}C_{0.08}N_{0.92}$). Micro-additions of other elements (Fe, Ni, Cr, and Mo), which participate in the formation of carbonitride precipitates result in small deviations from the equilibrium concentrations. We note that the initial composition is dependent on the concentrations of C and N in solid solution and temperature, both of which change the ratio $K_{(V,Nb)C}/K_{(V,Nb)N}$, because C and N (interstitial elements) have very large diffusivities compared to the substitutional elements V and Nb. Subsequent growth of these precipitates is limited by the diffusion of V and Nb from the matrix to the precipitates. In an ASB, the kinetics of carbonitride precipitation are, however, accelerated by a few-tens of orders of magnitude in the severely deformed austenite-phase [79]. Because VN is more stable thermodynamically than the other nitride and carbide phases, the vanadium and nitrogen are preferentially consumed from the matrix during the nucleation stage, while the carbon concentration increases by substituting for nitrogen at a later stage of growth [80]. This explains the enrichment of V and N in the core of the precipitates, and Nb and C in the outside shell of the growing precipitate, Fig. 7.

**4.3 Solute segregation and interaction effects on grain boundary migration in an adiabatic shear-band**

The results of the current APT experiments reveal a clear tendency for elemental segregation at low-angle sub-GBs and high-angle GBs. Most of the interstitial elements at sub-GBs (H, N, and C) are segregated during high strain-rate plastic deformation because they have large diffusivities in the medium-temperature range and interact readily with dislocations located in a sub-GB. In contrast, a high-angle GB also interacts strongly with not only interstitial elements but also with substitutional solute atoms (Mn, Ni, and Si). Prior APT research on steels has shown that elemental segregation occurs at both low and high-angle boundaries during heat treatment [42, 81, 82]. Severe deformation with a temperature increase within the ASB is, however, anticipated to redistribute these elements and reconfigure/eliminate these prior segregation profiles.

The interaction of solute atoms with a GB decreases its Gibbs interfacial free energy and the segregating solute atoms at a GB decreases the GB mobility during plastic deformation and recrystallization. The Gibbs interfacial free energy change of element *i* is described by the Gibbs' adsorption isotherm [83]:

$$d\gamma_i = -\Gamma_i \cdot d\mu_i \tag{8}$$

where $\mu_i$ is the chemical potential of element *i*, and $\Gamma_i$ is the Gibbsian excess free energy defined by eq. (1). In a dilute solid solution, the change of interfacial free energy due to the segregation of element i, $\Delta\gamma_i$, corresponds to values in excess of the initial concentration at these interfaces prior to segregation [84, 85]:

$$\Delta\gamma_i = -\Gamma_i \cdot k_B T \ln \frac{c_i}{c_o} \tag{9}$$

where $k_B$ is Boltzmann's constant, T is temperature in K, and $c_o$ and $c_i$ are the concentrations without and with segregation, respectively, at a GB. For the estimated peak temperature of 924 °C (section 4.1), the decreases in interfacial free energy due to segregation are -54.27± 4.92 mJ·m$^{-2}$ at GB-2 and - 40.86 ± 4.41 mJ·m$^{-2}$ at sub-GB-4, respectively. These values vary with the degree of segregation at an individual GB, which depends on the GB structure and its initial free energy [48, 50]. The solute-interactions with high-angle GBs and low-angle sub-GB provide information about how much additional energy is required to move an interface by shear-deformation.

Dislocation motion in a polycrystal can be restricted by various barriers, e.g., forcing dislocation pileups in local regions until they are released at a critical stress [86]. Thus, GBs are effective barriers to the movement of dislocations and the effective yield stress of a metal increases as the grain diameter decreases according to the Hall-Petch relationship [87, 88]. Solute-interactions at a GB or a sub-GB are regarded as the change of interfacial free energy, as shown in eq. (9). Segregation of solute atoms exerts a drag force on moving GBs and impedes their migration during deformation or recrystallization [89-91]. Thus, in an ASB, the energy stored by solute interactions with a GB (i.e. the extent of segregation) deters/obstructs shear-deformation or rotation of a sub-GB. This is analogous to solute-interactions with dislocations that deter their motion until at a critical break-away stress, an avalanche of dislocation motion occurs resulting in a stress drop and manifesting itself in the Portevin-Le Chatelier (PLC) effect in low-carbon steels [92]. A large value of the Gibbsian interfacial energy excess for a higher concentration of solute elements in an alloy increases the probability of localization and adiabatic shear band formation at a sufficiently high break-away shear stress, when significant GB migration and sub-GB rotation are triggered. This notion is supported by the fact that an interstitial-free steel does not exhibit the yield-point phenomenon and is only marginally prone to shear localization and ASB formation [93]. On the contrary, ASB formation becomes a dominant deformation mode in dynamic plastic deformation in nano-crystalline materials [94-96] and highly alloyed steels [97].

*4.4 Microstructural Evolution in adiabatic shear bands in a high-Ni steel*

As a final consideration, we discuss the formation mechanism of an ASB in a 10 wt.% Ni steel during high strain-rate deformation. The EBSD-TEM correlative studies describe the structural-transition from a lamellar morphology to equiaxed grains within a shear band. The breakdown of the highly sheared elongated lamellar structure to an equiaxed submicron-sized grain structure within the ASB is attributed to dynamic recovery and recrystallization processes [27, 64]. The EBSD orientation map (Fig. 4) demonstrates that the elongated lamellae contain many sub-grains (note color gradient) with a {101}, <111> texture along the shear direction (x-axis). The width of the individual lamella is reduced to 100 nm or less and finally it is fragmented in the vicinity of the transition line to yield small sub-grains by mechanical rotation. Recrystallization in the ASB is activated by local crystal rotations of sub-grains derived from a lamellar grain [98]. This is supported by the observation that fragmented grains from the lower side of the TL, Fig. 4, are grouped by a similar color in the orientation map because they are derived from one elongated lamellar grain. This group of grains evolves to an equiaxed random-orientation with limited grain growth towards the center of the shear band.

The classical kinetic models of recrystallization, GB migration, and sub-grain coalescence are inadequate to explain the observed small grain diameter in ASBs [61] because the time scale for these processes is several orders of magnitude larger than the deformation and cooling times experienced typically within an ASB. At high strain rates, the temperature increase with severe deformation in the ASB induces significant microstructure evolution, including dynamic recovery, recrystallization, and even amorphization in the extreme case. Mechanical deformation driven subgrain rotation has been proposed to account for the recrystallization observed in an ASB and is referred to as the *Progressive sub-grain misorientation* (*PrisM*) recrystallization model [24, 99]. The concept of accelerated dynamic recrystallization during fast deformation can be divided into three steps: (1) subgrain formation; (2) rotation/refinement by deformation; and (3) grain growth by sub-grain coalescence with further rotation and temperature increase.

Recently, Rittel et al. have suggested a different viewpoint in that dynamic localized shear failure of crystalline solids can be initiated by local dynamic recrystallization that occurs athermally, whose driving force is the stored energy of cold work rather than adiabatic heating and the associated temperature increase [100-103]. Once localization commences, then adiabatic heating plays a further role in modifying the microstructure. Thus, even though it is possible that the temperature effect is marginal in triggering dynamic recrystallization, the temperature rise accompanying shear localization can influence several microstructural events including grain growth in the core of the ASB, precipitation and dissolution, solute redistribution for segregation and other phase transformations, such as ferrite/martensite reverting to austenite.

Grain-rotation in the vicinity of the TL between the deformation-SB and transformation-SB is a critical step required for equiaxed grain formation in an ASB. Meyers et.al. [98] proposed a GB diffusion-assisted rotation kinetics model in a SB, Fig. 11 (a), where the mobility (M) for GB rotation is defined by $M = D_{gb}/k_B T$; where $k_B$ is the Boltzmann constant, T is temperature in Kelvin, and $D_{gb}$ is the GB-diffusivity. The relationship between the GB rotation angle ($\theta$), the time (t) for rotation of a grain at a temperature (T) is given by the following equation [24, 98],

$$t = \frac{kT \cdot L}{4\delta D_{gb} \gamma} \left( \frac{\tan\theta - 0.67\cos\theta}{1 - 2\sin\theta} + 0.77 \ln\left( \frac{\tan(\theta/2) - 2 - \sqrt{3}}{\tan(\theta/2) - 2 + \sqrt{3}} \right) - 1.361 \right) \qquad (10)$$

where $\gamma$ is the GB free energy, $\theta$ is the rotation angle of the grain, $\delta$ is the GB thickness, and L is the grain diameter. The maximum angle of rotation is assumed to be 30º as represented in Fig 11 (a). The GB energy ($\gamma$) in austenite is taken to be 0.65 J/m² Murr, 1975 #38794;Meyers, 2003 #37253} and the grain diameter (L = 0.1 µm in Fig. 4) for predicting the time (t) for rotation of a grain. Using values for the GB diffusivity and GB thickness (δ = 0.5 nm) given in the monograph by Frost and Ashby [104], the value of the time $\delta D_{gb}$ is given by $2\times 10^{-13} \exp(-Q_{GB}/RT)$, where $Q_{GB}$ is the activation energy for grain boundary diffusion ($Q_{GB}$ = 1.67 x10$^5$ J/mol [104]).

The predicted times for grain rotation in the range 0 ° up to 30 ° are displayed in Fig.11 (b) for the temperature range 800 to 1200 °C. The time ranges from 0.1 x 10$^{-6}$ to 4 x 10$^{-6}$ s in this temperature range, which is consistent with the deformation time for shear band formation. The deformation time (t) for a temperature increase in our specimen is 1.4×10$^{-4}$ s, which is obtained by dividing the total strain (0.4) by the strain-rate (2800 s$^{-1}$). Even if it is assumed that the last stage (5%) of deformation produces the temperature increase in the shear band, the time available is sufficient to induce grain rotation. The grain diameter (L) also influences the rotation, Eq. 11. At greater total strains or higher strain rates, the elongated grains subdivide into smaller sub-grains, until a critical grain diameter is achieved. At this point, to accommodate further deformation, the sub-grains rotate until they become highly misoriented equiaxed grains. This elongation and fragmentation of the lamellae, and their rotation to equiaxed grains agrees with our observations utilizing EBSD-TEM experiments.

In the final stage, *grain growth* occurs in the core of the ASB via subgrain coalescence by further rotation during adiabatic heating and subsequent cooling. From the TEM and EBSD analyses, the grains in the center of the ASB are larger than ones in the periphery close to the transition line and are equiaxed with a low density of imperfections. To evaluate the time frame for grain growth, the cooling time in the core of the ASB is presented in Fig. 11 (c). This calculated cooling curve is based on the error function heat equation [105]:

$$T = T_o/2 \left\{ erf \frac{\omega - h}{2\sqrt{\upsilon t}} + erf \frac{\omega + h}{2\sqrt{\upsilon t}} \right\} \qquad (11)$$

where $T_o$ is the highest temperature attained in the ASB, $\omega$ is the thickness of the shear band, $h$ is the is the length for heat trasport, $t$ is the time, and $\upsilon$ is the thermal diffusivity. The shear band thickness ($\omega$) was taken as ~21.7 μm, and was obtained from the SEM measurements in Fig. 2. The thermal diffusivity ($\upsilon$) is 3.78 mm$^2$/s, which is obtained from the thermal conductivity ($k$=13.4 W/m·K) divided by density (ρ =7.87 g/cm$^3$) and the specific heat capacity ($c_p$ =0.45 J/g·K), i.e., $\upsilon = k/\rho c_v$. Due to the small volume of the ASB and the large amount of surrounding material that acts as a heat sink, cooling is very fast: the time for cooling to <100 °C is less than $5 \times 10^{-3}$ s (dashed line in Fig. 11 (c)). The total heating and cooling time frame for the shear band in the sample tested at room temperature (i.e., ~ $1.4 \times 10^{-4}$ s deformation time plus $5 \times 10^{-3}$ s cooling time) is ~5.14 ms. This time frame is insufficient to induce classical diffusion-based recrystallization [61, 106]. From those heating and cooling times, it is concluded that grain growth is also driven by mechanical deformation within an ASB. The strong intermixing and temperature increase under high-strain-rate conditions in an ASB promotes grain boundary migration and subgrain coalescence to create a large degree of equiaxed grain with a low density of imperfections.

## 5. Summary and conclusions

A high-strength and high-toughness 10 wt.% Ni-steel has been developed for use of extrema work conditions due to their high strength and excellent crack-arrest toughness. The temperature increases and thermoplastic microstructural evolution in adiabatic shear bands were investigated utilizing correlative electron back-scatter diffraction (EBSD), transmission electron-microscopy (TEM), atom-probe tomography (APT) studies, from which the following conclusions are drawn:

- The uniaxial compression stress-strain curves, Fig. 1, display two different deformation responses at nominal strain rates of $10^{-4}$ s$^{-1}$ and 2800 s$^{-1}$. Under adiabatic-conditions, the temperature increases in the ASB were estimated by converting plastic work to heat-generation when the nominal height strain is $\geq 40\%$.
- The nominal structure of a dynamically deformed 10 wt.% Ni steel consists of lath martensite and two types of shear-bands, with or without phase-transformations or grain-rotations, Fig. 2. They are called transformed shear bands in the core and a deformed shear band on the periphery. Fine cracks were found along the transformed shear band, which can lead ultimately to a local decrease in the tensile strength and finally to fracture.
- TEM-EBSD analyses (Figs. 3,4) reveal that the deformed shear band has elongated lamellae with a shear texture {101}, <111> in the shear direction (SD). In contrast, the transformed shear band contains randomly oriented equiaxed grains ranging from 100-200 nm in diameter.

- The elongated grains in the deformed shear band are composed of a dislocation cell structure and subgrain boundaries. Rotation of subgrain boundaries within the elongated lamellar structure result in fragmented grains with a transition between the deformed- and transformed-bands, Fig. 11. This is described by the *progressive sub-grain misorientation model* (PrisM) and consists of three steps: (1) subgrain formation; (2) rotation/refinement by deformation; and (3) grain growth by subgrain coalescence with further rotation.
- In Fig. 5, the alloying elements, Ni, Mn, C, N(Si), and H, segregate at grain- boundaries (GBs) in the deformed shear band with Gibbsian interfacial excess values, $\Gamma_i$, of 1.65 ± 0.43, 0.20 ± 0.13, 1.88 ± 0.12, 2.56 ± 0.13, and 1.07 ± 0.15 atom·nm$^{-2}$, respectively. Inside the grains, the subgrain boundaries can be identified by the strong segregation of interstitial elements, N (or Si), H, and C with Gibbsian interfacial excess values, $\Gamma_i$, of 0.40 ± 0.12, 1.73 ± 0.16, 0.11± 0.15, and 1.02 ± 0.25 atom·nm$^{-2}$, respectively.
- The decreases in interfacial free energies due to segregation at the peak temperature of 924 °C are -54.27 ± 4.92 mJ·m$^{-2}$ at the GBs and - 40.86 ± 4.41mJ·m$^{-2}$ at the subgrain boundaries, respectively. The solute atom interactions with high-angle GBs and sub-grain boundaries provide the necessary decrease in interfacial free energy required to move these interfaces by shear deformation. The solute atom interactions can increase the instability of localized shear deformation by promoting depinning of the coupled grain boundary boundaries.
- The chemical compositions in the transformed shear band are redistributed by temperature increases and an austenite phase transformation. The TEM-APT analyses, Fig. 6, demonstrate the enrichment of carbon and other austenite stabilizing elements with the following partitioning ratios ($K_i^{\gamma/\alpha}$): 5.09 ±0.27 (C), 1.30 ±0.06 (N and Si), 1.08 ± 0.19 (Ni), 1.38 ± 0.07 (Mn), respectively. The interstitial elements, especially carbon, play an important role in stabilizing austenite that forms during adiabatic heating, particularly given the extremely short duration of the temperature excursion ($\sim 1.4\times10^{-4}$ s).
- V(Nb) carbonitride precipitates, 3-7 nm in diam and with a number density of 2.2 ± 0.6 x 10$^{23}$/m$^3$, were observed in the transformed shear band, Fig. 7. The precipitates are composed of 40.16 ± 1.45 V and 34.51 ± 0.94 N, to 7.65 ± 0.76 Nb, 2.67 ± 0.46 C, 4.12 ± 0.30 Cr, and 0.77 ± 0.25 Mo at.%. The precipitate composition, Nb$_x$V$_{1-x}$C$_y$N$_{1-y}$, was estimated by a quasi-regular-solution thermodynamic model as a function of temperature, Fig. 10. The temperature for complete dissolution of the precipitates is ~1058 ºC, and the calculated composition is $Nb_{0.17}V_{0.83}C_{0.12}N_{0.88}$ (x = 0.17 and y = 0.12) at 924 ºC. These values are in reasonable agreement with the APT experimental result for the composition of V(Nb) carbonitride-precipitate ($Nb_{0.16}V_{0.84}C_{0.08}N_{0.92}$).

**Acknowledgements**

The authors (S-IB and DNS) gratefully acknowledge financial support for this research by the Office of Naval Research through Grant Number N000141812594 and useful discussions with the Program Manager, Dr. William Mullins, of the Office of Naval Research. RKG and KSK thank the Office of Naval Research (Contract # N00014-05-1-0062) for financial support with Dr. J. Christodoulou as the Program Manager and Dr. X.J. Zhang of NSWCCD for providing the material and for several useful discussions. The authors also thank Mr. Ding-Wen (Tony) Chung for helpful discussions about the thermodynamic composition of precipitates. Atom-probe tomography was performed at the Northwestern University Center for Atom-Probe Tomography (NUCAPT). The LEAP tomograph at NUCAPT was purchased and upgraded with grants from the NSF-MRI (DMR-0420532) and ONR-DURIP (N00014-0400798, N00014-0610539, N00014-0910781, N00014-1712870) programs. NUCAPT received support from the MRSEC program (NSF DMR-1720139) at the Materials Research Center, the SHyNE Resource (NSF ECCS-1542205), and the Initiative for Sustainability and Energy (ISEN) at Northwestern University. This research also made use of instruments in Northwestern's NUANCE-EPIC. NUCAPT and EPIC received support from the MRSEC program (NSF DMR-1121262) through Northwestern's Materials Research Center. EPIC received support from the International Institute for Nanotechnology (IIN) and the State of Illinois, through the IIN. We thank research associate professor Dieter Isheim for his management of NUCAPT and useful discussions.

## Table captions

**Table 1.** APT measured compositions of ferrite matrix, maximum compositions at GB2 and sub-GB4, and their interfacial excesses in the deformation shear band region. GB2 and sub-GB4 were selected to measure the highest position of composition in the proximity histogram and interfacial excesses. (at.%)

| ELEMENT | Ni | Mn | C | N, Si | H | Fe |
|---|---|---|---|---|---|---|
| Ferrite (at.%) | 10.62 ± 0.16 | 0.76 ± 0.10 | 0.22 ± 005 | 0.54 ± 0.09 | 0.42 ± 0.05 | 84.42 ± 0.25 |
| Max. at GB2 (at.%) | 10.72 ± 0.23 | 0.82 ± 0.7 | 1.07 ± 0.07 | 1.47 ± 0.09 | 0.90 ± 0.7 | 82.42 ± 0.28 |
| $\Gamma_i^{GB2}$ | 1.65 ± 0.43 | 0.20 ± 0.13 | 1.88 ± 0.12 | 2.56 ± 0.13 | 1.07 ± 0.15 | -6.95 ± 0.51 |
| Max. at Sub-GB4 (at.%) | 10.98 ± 0.38 | 0.78 ± 0.21 | 0.42 ± 0.07 | 1.78 ± 0.15 | 0.85 ± 0.15 | 82.38 ± 0.46 |
| $\Gamma_i^{S-GB4}$ | 1.02 ± 0.58 | 0.11 ± 0.15 | 0.40 ± 0.12 | 2.29 ± 0.15 | 1.08 ± 0.25 | -5.28 ± 0.46 |

**Table 2.** Compositions (at.%) of ferrite matrix, austenite, vanadium-rich carbonitride precipitates, and their partitioning ratios (K) values in a transformed shear band.

| Element | Ni | Mn | C | N, Si | Mo | Cr | V | Nb | Fe |
|---|---|---|---|---|---|---|---|---|---|
| Ferrite | 10.62 ± 0.16 | 0.76 ± 0.10 | 0.22 ± 0.05 | 0.54 ± 0.09 | 0.66 ± 0.02 | 0.67 ± 0.02 | 0.12 ± 0.01 | 0.02 ± 0.01 | 84.42 ± 0.25 |
| Austenite | 11.48 ± 0.16 | 1.05 ± 0.05 | 1.12 ± 0.06 | 0.70 ± 0.04 | 0.62 ± 0.04 | 0.69 ± 0.04 | 0.38 ± 0.09 | 0.05 ± 0.03 | 82.57 ± 0.21 |
| Aust./Fer. partitioning ($K_i^{\gamma/\alpha}$) | 1.08 ± 0.19 | 1.38 ± 0.07 | 5.09 ± 0.27 | 1.30 ± 0.06 | 0.94 ± 0.04 | 1.03 ± 0.05 | 3.17 ± 0.16 | 2.5 ± 0.09 | 0.98 ± 0.25 |
| Precipitate | 1.17 ± 0.26 | 0.31 ± 0.12 | 2.67 ± 0.46 | 34.51± 0.94 | 0.77 ± 0.25 | 4.12 ± 0.30 | 40.16 ± 1.45 | 7.65 ± 0.76 | 6.57 ± 0.68 |
| Precip./Fer. partitioning ($K_i^{p/\alpha}$) | 0.11 ± 0.02 | 0.41 ± 0.16 | 12.0 ± 0.71 | 63.91 ± 2.94 | 1.17 ± 0.38 | 6.15 ± 0.48 | 334.7 ± 30.4 | 382.5 ± 84.2 | 0.08 ± 0.01 |

**Table 3.** Solubility products ($K_s$) in the Henrian form, $\log_{10} K_s = \log [M][X] = A - B/T$

| Compound | A | B | reference |
|----------|------|-------|-----------|
| NbC | 2.26 | 6770 | [75] |
| NbN | 4.04 | 10230 | [76] |
| VC | 6.72 | 9500 | [77] |
| VN | 3.02 | 7840 | [77] |

# Figure captions

**Fig. 1.** True stress-strain curves obtained from a conventional uniaxial compression test at the nominal quasi-static strain-rate of $10^{-4}$ s$^{-1}$ and from a split-Hopkinson bar set-up at a high strain-rate of 2800 s$^{-1}$.

**Fig. 2.** Scanning electron microscopy (SEM) image of an adiabatic shear band (ASB). (a) Transformed shear band in the middle of the ASB with the likelihood of a phase transformation or grain rotation and a thickness of ~ 3.3 ± 0.5 µm. Outside the transformed band, the striations are closely-spaced and aligned in the shear direction, indicating extensive plastic flow (marked as deformed shear band with a thickness of ~21.7 ± 1.5 µm). (b) Fine microcracks are present in the shear band, an example of which is displayed. Plan view TEM/APT sample from the shear band was obtained from the region indicated by the white rectangle.

**Fig. 3.** Plan view transmission electron microscopy (TEM) images of an adiabatic shear band (ASB) including a crack tip; the location-specific specimen was lifted-out utilizing focused ion beam (FIB) microscopy. (a) Low magnification image of the entire sample; (b) bright field (BF) image of the transition from the elongated deformed grain to the equiaxed grains across the dashed transition line (TL), (c) Center region of the ASB with the crack tip; (d) High magnification image of the secondary microcrack region within the ASB, ahead of the main crack; (e, f) Dark field (DF) images corresponding to (b, c) obtained from the (110) spots indicated by the circles in the selected area diffraction pattern (SADP) in the upper right inset.

**Fig. 4.** Electron back scatter diffraction (EBSD) orientation map across the transition region from the deformed to the transformed shear band in the ASB. (a) Inverse pole figure (IPF) map across the transition line of an ASB. The reference frame of the orientation map is represented by the shear direction (SD, X-axis) and the shear plane normal (SPN, Y-axis). Small angle of misorientations (<1.5 degrees) and twin boundaries (<111>, 60, {112}) within the grain are represented as gray lines and white lines, respectively. The transition from elongated lamellar grains to equiaxed small grains is represented by the blue dotted transition line (TL). (b) Pole figure maps derived from (a). Shear texture {101}, <111> can be observed in the deformed shear band (Y- and X-axes in pole figure map). This implies that <111>-direction of the b.c.c. crystal is parallel to the shear-direction (X-axis) and the normal to the {110} plane is parallel to the shear-plane normal (Y-axis).

**Fig. 5.** Correlative TEM and 3-D APT analyses in the deformed shear band obtained ~ 3 µm away from the transition line. (a) A BF TEM image of the APT nanotip with a SADP in the grain between GB-3 and GB-4. The SADP is near the [112] zone axis of b.c.c.-iron. (b, c) Two 3-D APT reconstructed images with ~ 133 and ~120 million atoms, respectively. Concentration profiles across (d) GBs (1, 2, 4, 5) and (e) sub-GBs (SG1 - S4), demonstrating segregation. The GBs can be identified in the APT images by comparing the profiles of the elemental segregation in the 3-D APT reconstruction with the GB contrast in the TEM image (a). Inside the grain, the sub-GBs are detected from N (or Si) segregation (green dots). The N and Si peaks in a time-of-flight (TOF) mass-to-charge-state-ratio spectrum overlap with one another in an APT experiment. The segregation behavior of a solute species, $i$, at GBs and sub-GBs are described by Gibbsian interfacial excesses, $\Gamma_i$ [83] as presented in Table 1.

**Fig. 6.** Correlative TEM and the 3-D APT analyses in the transformed shear band at the core of the ASB. (a) A BF TEM image of an APT nanotip containing six grains exhibiting different diffraction contrast effects. (b) 3-D APT reconstructed image containing ~133 million atoms. Only the carbon atoms (black dots) are displayed for clarity; the fully colored image is displayed in Fig. 7 (a2). (c) Concentration profiles obtained along the z-axis of the 3-D APT reconstruction in Fig 6 (b). The carbon concentration increases to a maximum value of 1.5 at.%, with an average concentration of $1.12 \pm 0.06$ at.% in the austenite grain. The elements N(Si), Mn, and Ni are also enriched within the austenite grain, with values of $0.70 \pm 0.06$, $1.05 \pm 0.06$, and $11.48 \pm 0.16$ at.%, respectively. The average concentrations and partitioning ratios between the ferrite matrix and the austenite phase are presented in Table 2.

**Fig. 7.** 3-D APT analyses of vanadium-rich carbonitride precipitates in the core of the ASB. (a) 3-D reconstructed image of an APT nanotip with carbonitride precipitates employing a 20 at.% vanadium iso-concentration surface. (b) Proximity histogram concentration profiles for Fe, Ni, V, N, Nb, Cr, C, Mn, and Mo, with respect to distance from the matrix/precipitate interface, the origin (0). The vertical black dotted line is through the inflection point of the Fe concentration profile. The average concentrations and partitioning ratios between the ferrite matrix and precipitates are presented in Table 2.

**Fig. 8.** Computed stress-strain curves for isothermal conditions at strain rates of $10^{-4}$ s$^{-1}$ (black solid black) and 2800 s$^{-1}$ (blue dashed) and after incorporating adiabatic conditions (red dashed dots), assuming for the latter an initial temperature of 300 K.

**Fig. 9.** Temperature increase with strain, eq. (2), at a nominal strain rate of 2800 s$^{-1}$ for homogeneous deformation (solid blue) and localized deformation from initial strain (red dots). The calculations for the onset of localization onset from the very beginning demonstrates that the temperature can achieve 924 °C, when the magnitude of the strain approaches 40%. The dashed green line displays the case when the shear localization commences at ~ 35% strain with a high localization factor ($V_o/V_p$) of 50.

**Fig. 10.** The compositional fractions (x, y) and the mole fraction (f) of (Nb, V) carbonitride, Nb$_x$V$_{1-x}$C$_y$N$_{1-y}$ precipitates. The temperature for complete dissolution is ~1058 °C. At an estimated temperature of 924 °C, the composition of the carbonitride is Nb$_{0.17}$V$_{0.83}$C$_{0.12}$N$_{0.88}$ (x = 0.17 and y = 0.12) and the mole fraction (f) is 5.3x10$^{-4}$, values that are in agreement with the experimental APT results (Nb$_{0.16}$V$_{0.84}$C$_{0.08}$N$_{0.92}$). When the temperature decreases from 924 °C, the proportion of carbon in the precipitates increases but the Nb fraction decreases.

**Fig. 11.** Dynamic recrystallization with subgrain rotation accelerated by high strain-rate deformation and a temperature increase within an ASB. (a) Model for grain rotation kinetics for dynamic recrystallization [98]; (b) predicted times for grain rotation in the temperature range, 800-1200 °C. The time ranges from 0.1 to 4 microseconds in the temperature regime considered, which is reasonably consistent with the deformation time for shear band formation; and (c) Cooling curve as a function of time at the core of the ASB. The initial heating temperature ($T_o$) was set at 924 °C.

**Figure 1**

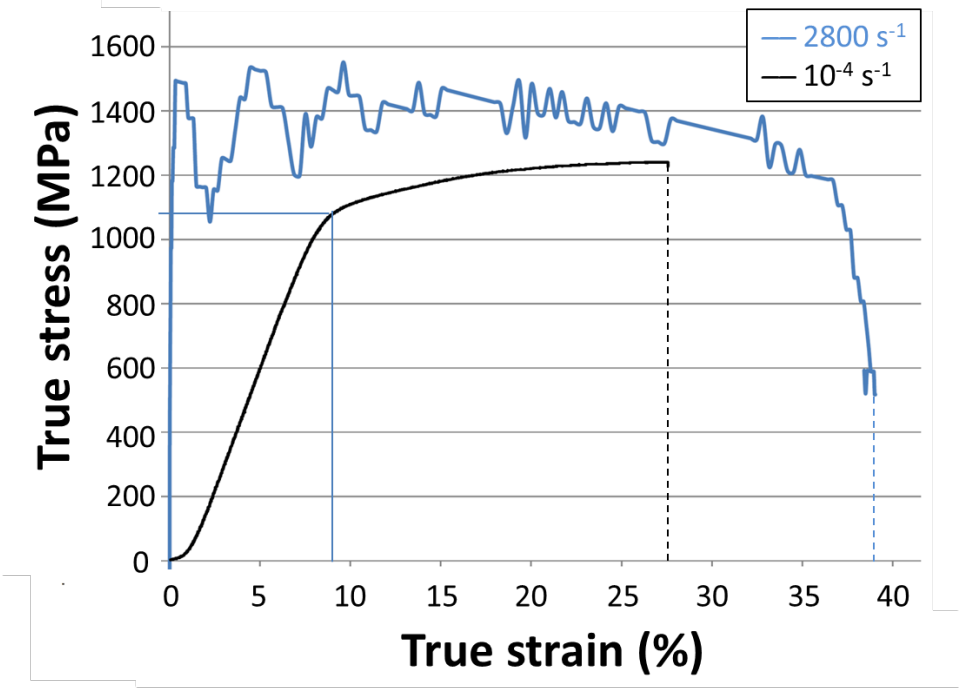

**Fig. 1.** True stress-strain curves obtained from a conventional uniaxial compression test at the nominal quasi-static strain-rate of $10^{-4}$ s$^{-1}$ and from a split-Hopkinson bar set-up at a high strain-rate of 2800 s$^{-1}$.

**Figure 2**

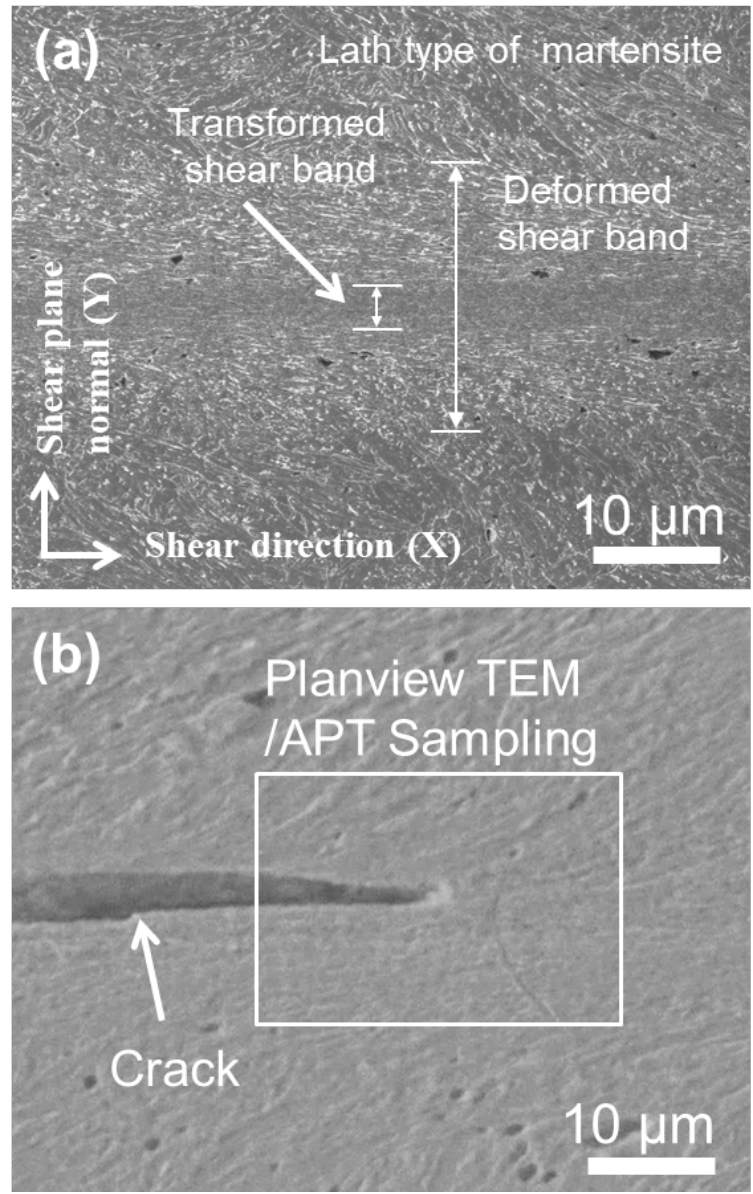

**Fig. 2.** Scanning electron microscopy (SEM) image of an adiabatic shear band (ASB). (a) Transformed shear band in the middle of the ASB with the likelihood of a phase transformation or grain rotation and a thickness of ~ 3.3 ± 0.5 µm. Outside the transformed band, the striations are closely-spaced and aligned in the shear direction, indicating extensive plastic flow (marked as deformed shear band with a thickness of ~21.7 ± 1.5 µm). (b) Fine microcracks are present in the shear band, an example of which is displayed. Plan view TEM/APT sample from the shear band was obtained from the region indicated by the white rectangle.

**Figure 2**

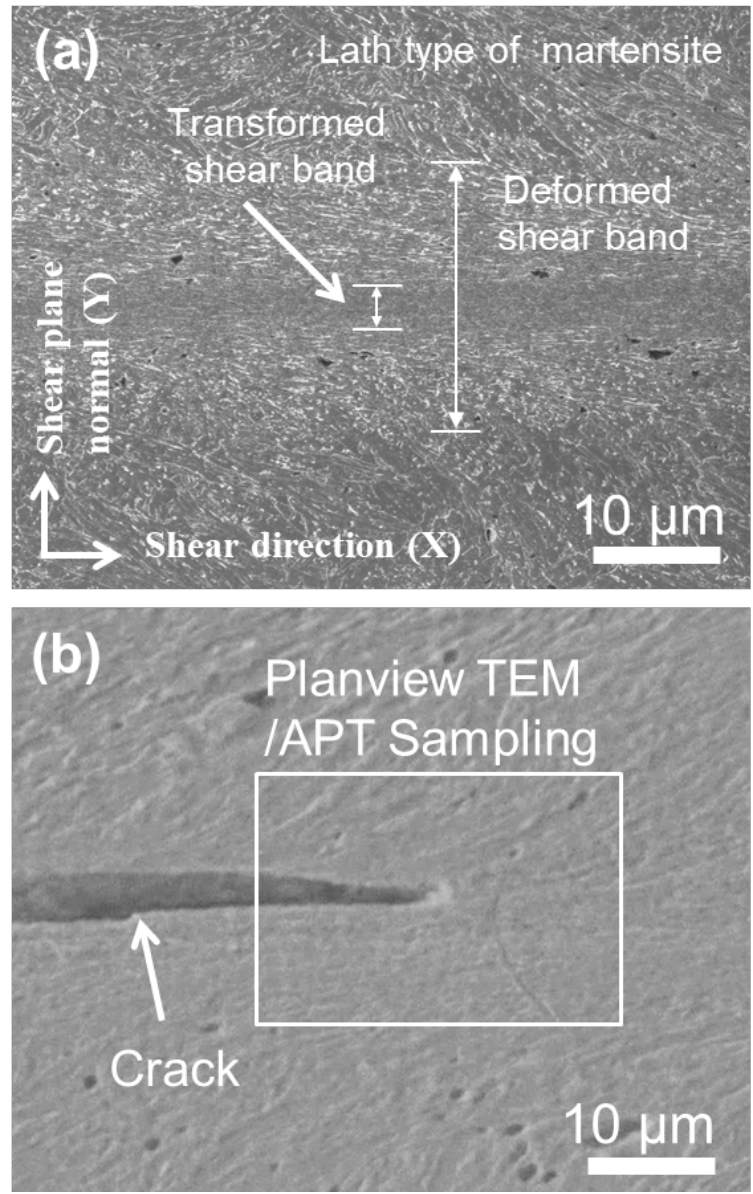

**Fig. 2.** Scanning electron microscopy (SEM) image of an adiabatic shear band (ASB). (a) Transformed shear band in the middle of the ASB with the likelihood of a phase transformation or grain rotation and a thickness of ~ 3.3 ± 0.5 µm. Outside the transformed band, the striations are closely-spaced and aligned in the shear direction, indicating extensive plastic flow (marked as deformed shear band with a thickness of ~21.7 ± 1.5 µm). (b) Fine microcracks are present in the shear band, an example of which is displayed. Plan view TEM/APT sample from the shear band was obtained from the region indicated by the white rectangle.

*Figure 3*

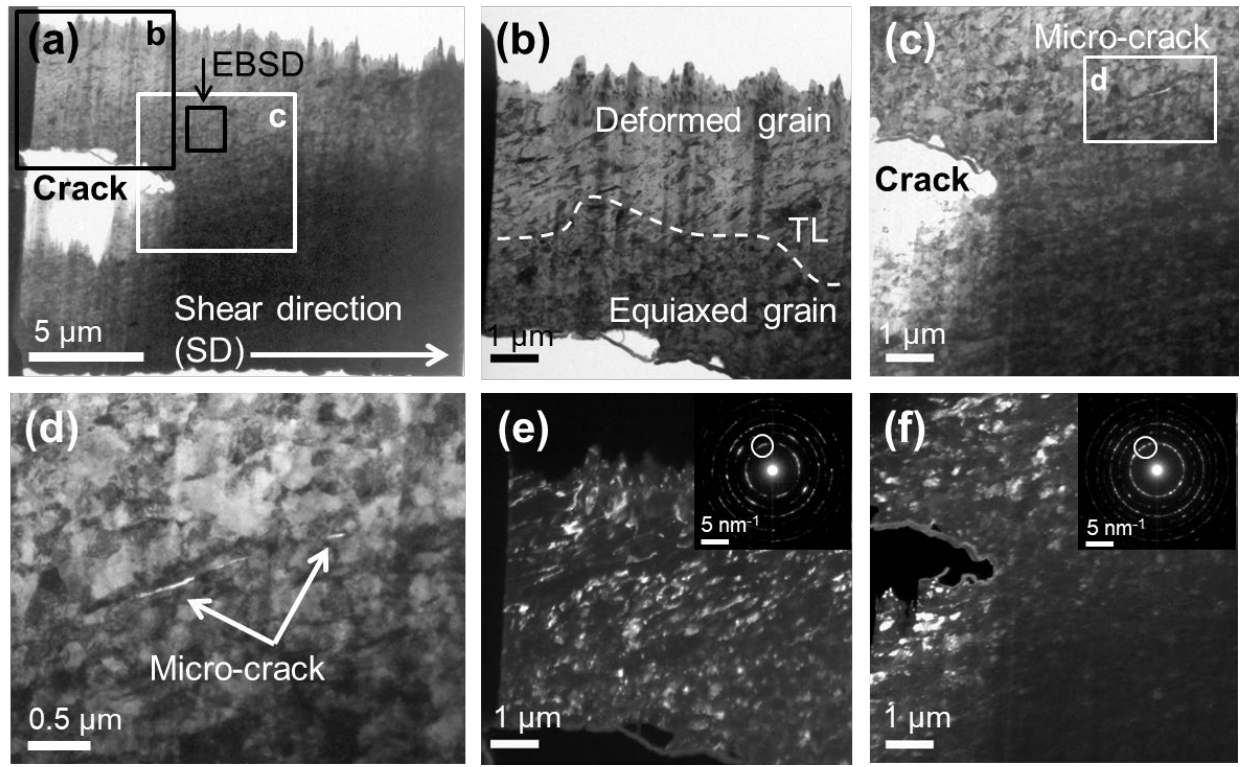

**Fig. 3.** Plan view transmission electron microscopy (TEM) images of an adiabatic shear band (ASB) including a crack tip; the location-specific specimen was lifted-out utilizing focused ion beam (FIB) microscopy. (a) Low magnification image of the entire sample; (b) bright field (BF) image of the transition from the elongated deformed grain to the equiaxed grains across the dashed transition line (TL), (c) Center region of the ASB with the crack tip; (d) High magnification image of the secondary microcrack region within the ASB, ahead of the main crack; (e, f) Dark field (DF) images corresponding to (b, c) obtained from the (110) spots indicated by the circles in the selected area diffraction pattern (SADP) in the upper right inset.

Figure 4

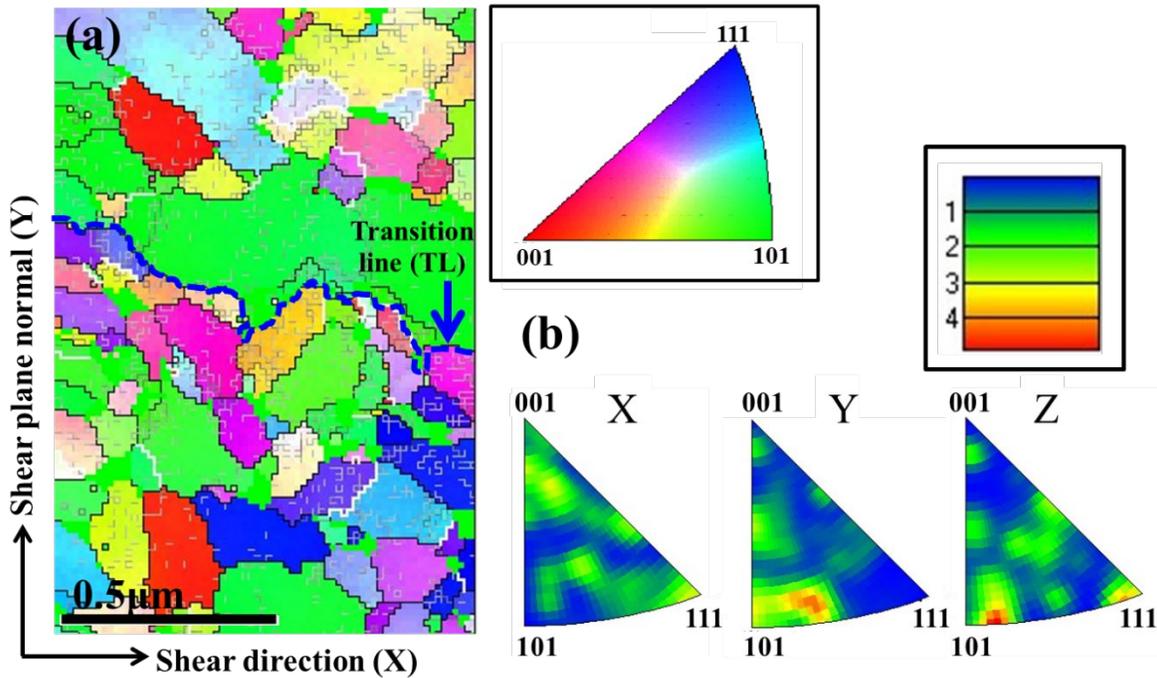

**Fig. 4.** Electron back scatter diffraction (EBSD) orientation map across the transition region from the deformed to the transformed shear band in the ASB. (a) Inverse pole figure (IPF) map across the transition line of an ASB. The reference frame of the orientation map is represented by the shear direction (SD, X-axis) and the shear plane normal (SPN, Y-axis). Small angle of misorientation (<1.5 degrees) and twin boundaries (<111>, 60, {112}) within the grain are represented as gray lines and white lines, respectively. The transition from elongated lamellar grains to equiaxed small grains is represented by the blue dotted transition line (TL). (b) Pole figure maps derived from (a). Shear texture {101}, <111> can be observed in the deformed shear band (Y- and X-axes in pole figure map). This implies that <111>-direction of the b.c.c. crystal is parallel to the shear-direction (X-axis) and the normal to the {110} plane is parallel to the shear-plane normal (Y-axis).

*Figure 5*

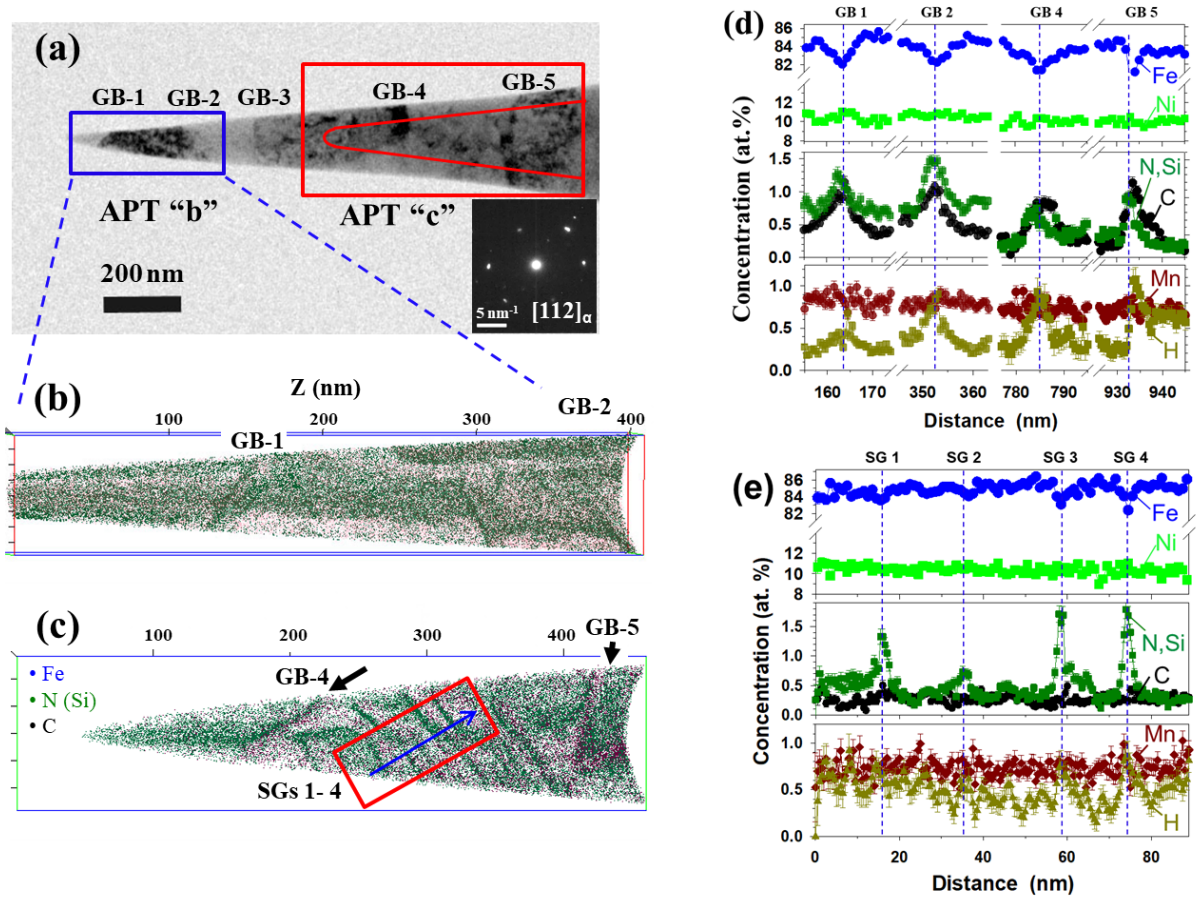

**Fig. 5.** Correlative TEM and 3-D APT analyses in the deformed shear band obtained ~ 3 μm away from the transition line. (a) A BF TEM image of the APT nanotip with a SADP in the grain between GB-3 and GB-4. The SADP is near the [112] zone axis of b.c.c.-iron. (b, c) Two 3-D APT reconstructed images with ~ 133 and ~120 million atoms, respectively. Concentration profiles across (d) GBs (1, 2, 4, 5) and (e) sub-GBs (SG1 - S4), demonstrating segregation. The GBs can be identified in the APT images by comparing the profiles of the elemental segregation in the 3-D APT reconstruction with the GB contrast in the TEM image (a). Inside the grain, the sub-GBs are detected from N (or Si) segregation (green dots). The N and Si peaks in a time-of-flight (TOF) mass-to-charge-state-ratio spectrum overlap with one another in an APT experiment. The segregation behavior of a solute species, *i*, at GBs and sub-GBs are described by Gibbsian interfacial excesses, $\Gamma_i$ [83] as presented in Table 1.

Figure 6

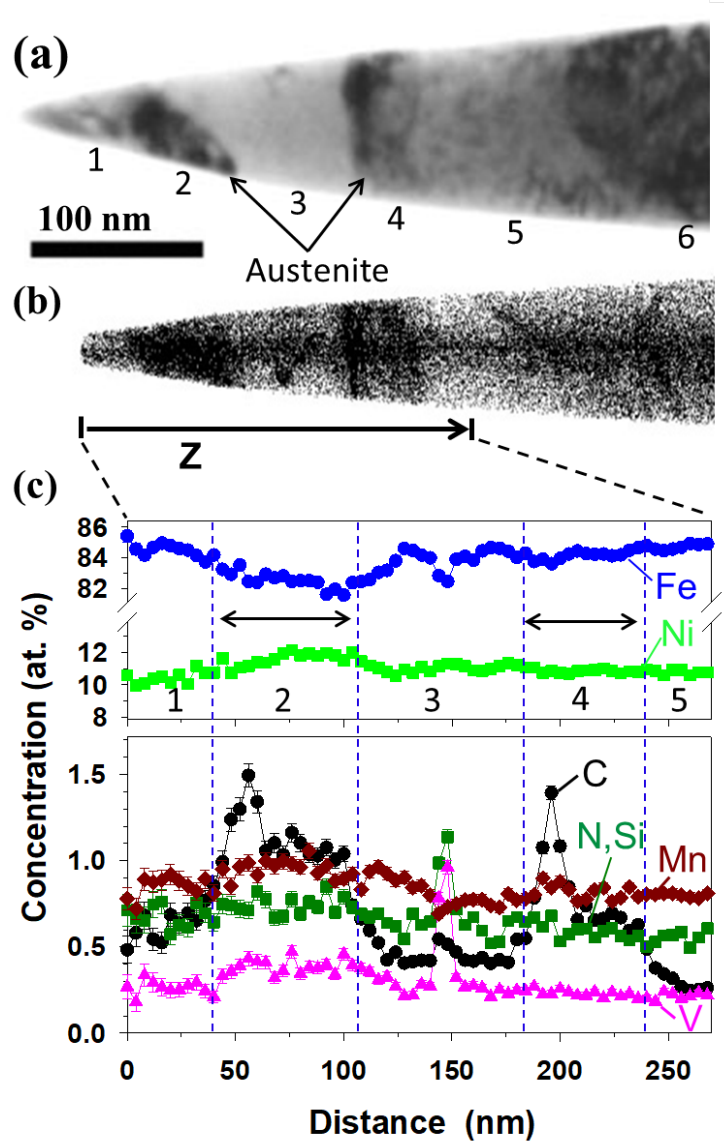

**Fig. 6.** Correlative TEM and the 3-D APT analyses in the transformed shear band at the core of the ASB. (a) A BF TEM image of an APT nanotip containing six grains exhibiting different diffraction contrast effects. (b) 3-D APT reconstructed image containing ~133 million ions. Only the carbon atoms (black dots) are displayed for clarity; the fully colored image is displayed in Fig. 7 (a2). (c) Concentration profiles obtained along the z-axis of the 3-D APT reconstruction in Fig 6 (b). The carbon concentration increases to a maximum value of 1.5 at.%, with an average concentration of $1.12 \pm 0.06$ at.% in the austenite grain. The elements N(Si), Mn, and Ni are also enriched within the austenite grain, with values of $0.70 \pm 0.06$, $1.05 \pm 0.06$, and $11.48 \pm 0.16$ at.%, respectively. The average concentrations and partitioning ratios between the ferrite matrix and the austenite phase are presented in Table 2.

**Figure 7**

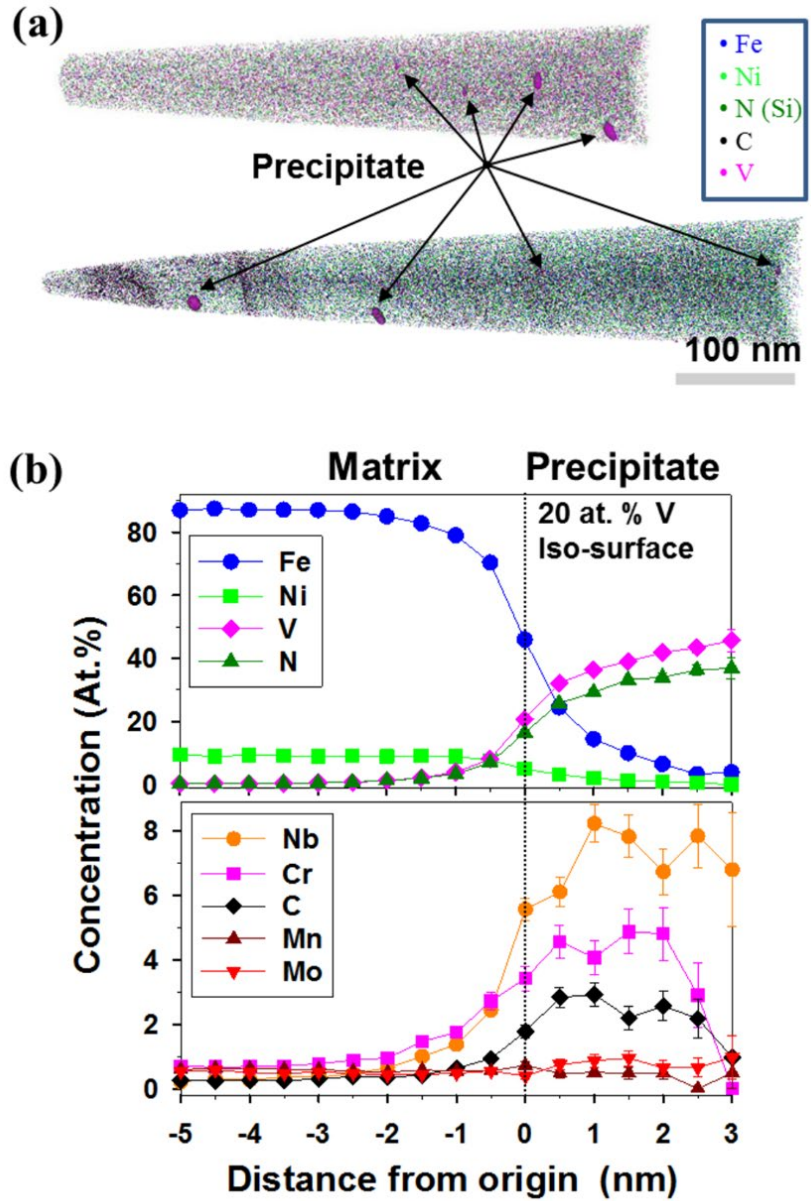

**Fig. 7.** 3-D APT analyses of vanadium-rich carbonitride precipitates in the core of the ASB. (a) 3-D reconstructed image of an APT nanotip with carbonitride precipitates employing a 20 at.% vanadium iso-concentration surface. (b) Proximity histogram concentration profiles for Fe, Ni, V, N, Nb, Cr, C, Mn, and Mo, with respect to distance from the matrix/precipitate interface, the origin (0). The vertical black dotted line is through the inflection point of the Fe concentration profile. The average concentrations and partitioning ratios between the ferrite matrix and precipitates are presented in Table 2.

*Figure 8*

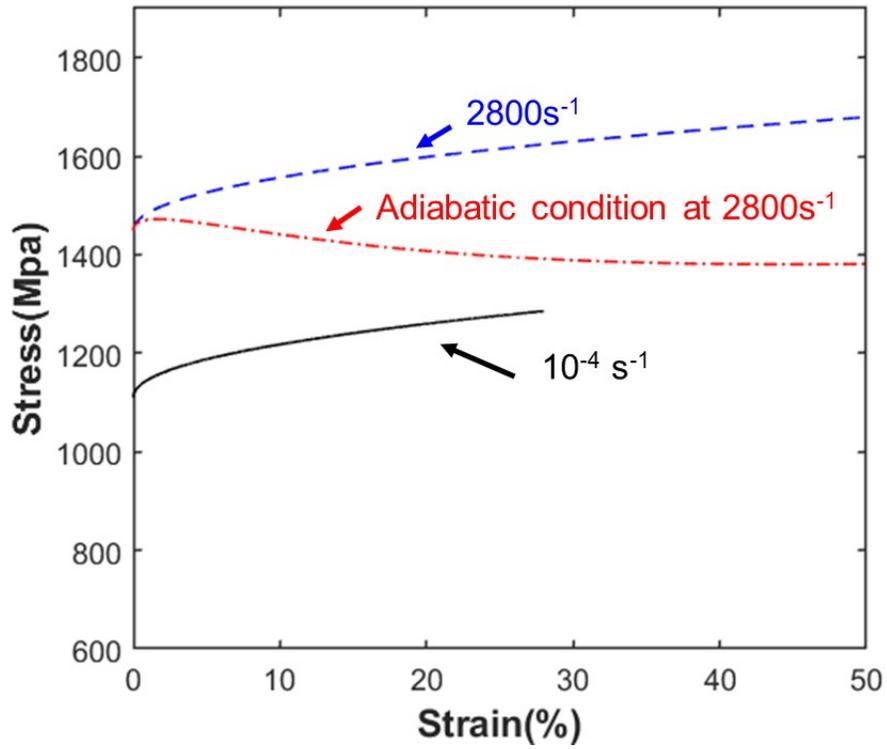

**Fig. 8.** Computed stress-strain curves for isothermal conditions at strain rates of $10^{-4}$ s$^{-1}$ (black solid black) and 2800 s$^{-1}$ (blue dashed) and after incorporating adiabatic conditions (red dashed dots), assuming for the latter an initial temperature of 300 K.

**Figure 9**

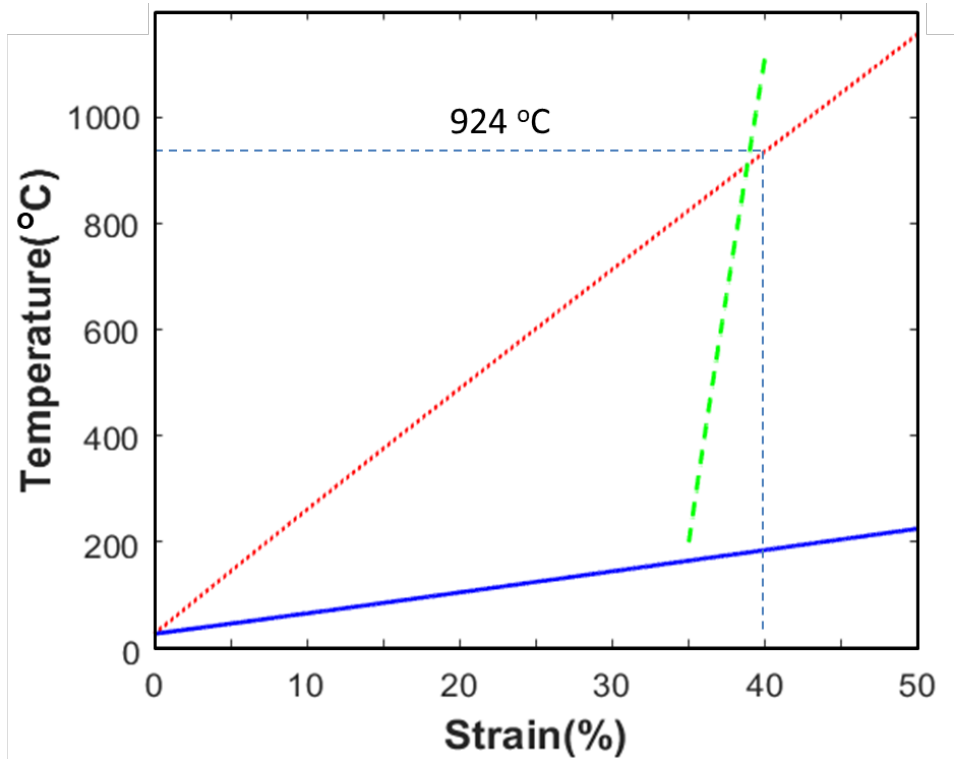

**Fig. 9.** Temperature increase with strain, eq. (2), at a nominal strain rate of 2800 s$^{-1}$ for homogeneous deformation (solid blue) and localized deformation from initial strain (red dots). The calculations for the onset of localization onset from the very beginning demonstrates that the temperature can achieve 924 °C, when the magnitude of the strain approaches 40%. The dashed green line displays the case when the shear localization commences at ~ 35% strain with a high localization factor ($V_o/V_p$) of 50.

**Figure 10**

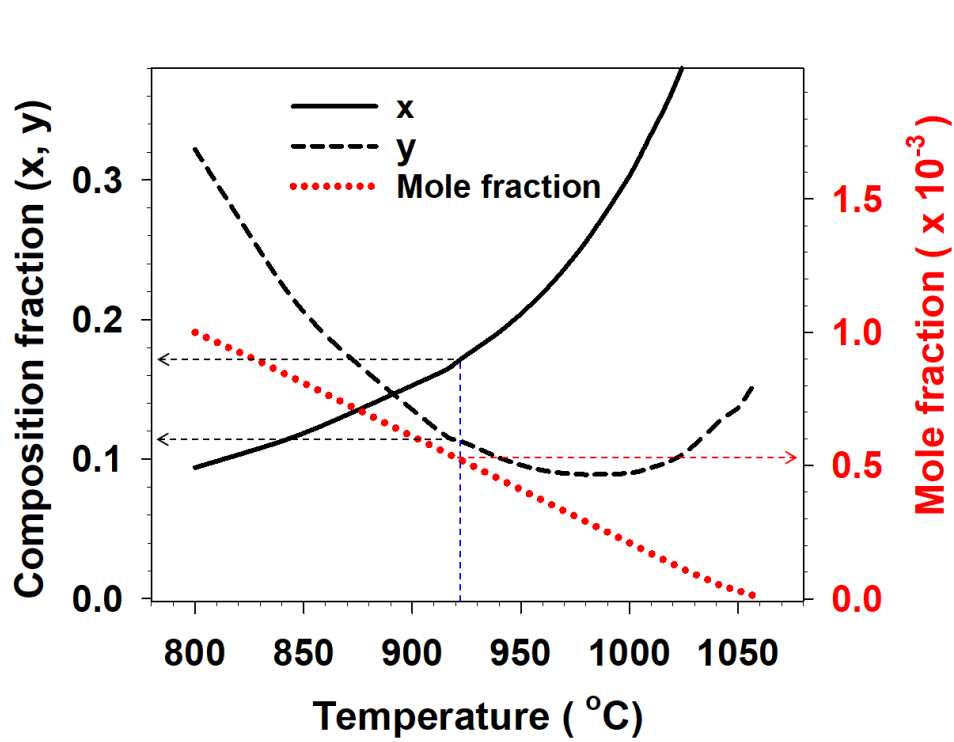

**Fig. 10.** The compositional fractions (x, y) and the mole fraction (f) of (Nb, V) carbonitride, $Nb_xV_{1-x}C_yN_{1-y}$ precipitates. The temperature for complete dissolution is ~1058 °C. At an estimated temperature of 924 °C, the composition of the carbonitride is $Nb_{0.17}V_{0.83}C_{0.12}N_{0.88}$ (x = 0.17 and y = 0.12) and the mole fraction (f) is $5.3 \times 10^{-4}$, values that are in agreement with the experimental APT results ($Nb_{0.16}V_{0.84}C_{0.08}N_{0.92}$). When the temperature decreases from 924 °C, the proportion of carbon in the precipitates increases but the Nb fraction decreases.

Figure 11

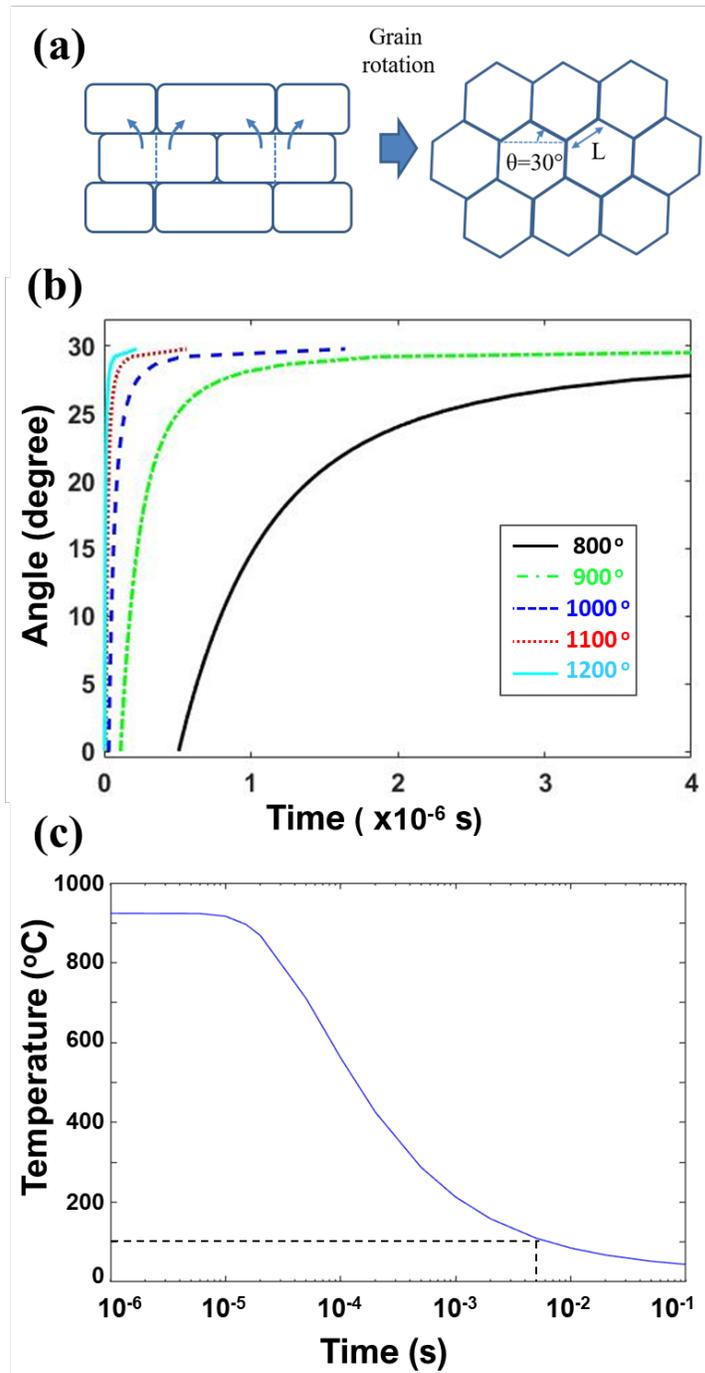

**Fig. 11.** Dynamic recrystallization with subgrain rotation accelerated by high strain-rate deformation and a temperature increase within an ASB. (a) Model for grain rotation kinetics for dynamic recrystallization [98]; (b) predicted times for grain rotation in the temperature range, 800-1200 °C. The time ranges from 0.1 to 4 microseconds in the temperature regime considered, which is reasonably consistent with the deformation time for shear band formation; and (c) Cooling curve as a function of time at the core of the ASB. The initial heating temperature ($T_o$) was set at 924 ºC.